# The 10-point and 12-point Number Theoretic Hilbert Transform

Vamsi Sashank Kotagiri
Oklahoma State University, Stillwater

**Abstract**
This paper presents 10-point and 12-point versions of the recently introduced number theoretic Hilbert (NHT) transforms. Such transforms have applications in signal processing and scrambling. Polymorphic solutions with respect to different moduli for each of the two cases have been found. The multiplicity of solutions for the same moduli increases their applicability to cryptography.

**Introduction**
The NHT, introduced recently [1] as a generalization of the discrete Hilbert transform (DHT) [2], is a circulant matrix with alternating entries of each row being zero and non-zero numbers and transpose modulo a suitable number is its inverse. The DHT has found many applications in signal processing and also in scrambling [3]-[5] and in a variety of other applications in speech and image analysis [6]-[12]. The relevant background information on number theoretic transforms in provided in [13]-[17]. One would expect, therefore, that NHT will also have many applications.

In the earlier paper [2], only 4-point, 6-point, and 8-point NHT matrices were presented. Here we consider examples of 10-point and 12-point NHT which provide us greater flexibility in their use.

Let us consider the data block to be F and the NHT transform to be N. The matrix N is the general form of NHT and m is appropriate value of modulus; its inverse is $N^T$ mod m.

$$N = \begin{bmatrix} 0 & a & 0 & b & 0 & c & 0 & . & . & k \\ k & 0 & a & 0 & b & 0 & c & 0 & . & . \\ . & k & 0 & a & 0 & b & 0 & c & 0 & . \\ . & . & k & 0 & a & 0 & b & 0 & c & 0 \\ 0 & . & . & k & 0 & a & 0 & b & 0 & c \\ c & 0 & . & . & k & 0 & a & 0 & b & 0 \\ 0 & c & 0 & . & . & k & 0 & a & 0 & b \\ b & 0 & c & 0 & . & . & k & 0 & a & 0 \\ 0 & b & 0 & c & 0 & . & . & k & 0 & a \\ a & 0 & b & 0 & c & 0 & . & . & k & 0 \end{bmatrix} \mod m$$

For a block of data F, the NHT transform $G = NF$ mod $m$. The inverse of the transform is

$$F = N^T G \bmod m$$

**10-point NHT**

The 10-point NHT matrix be given by the following matrix and the first row of the 10-point NHT is 0,a,0,b,0,c,0,d,0,e. If we multiply the 10-point NHT with its transpose we observe that

$$a^2 + b^2 + c^2 + d^2 + e^2$$

is the diagonal element term and the non-diagonal element terms are

(b+ e) a + (c+ e) d + bc and (a+ e) c +(b + a)d + eb.

In order to get a valid NHT matrix we need to assume a suitable modulus in such that all the non-diagonal elements of the matrix product $NN^T$ will become zero and only the diagonal elements of the product matrix remain. There are many solutions which satisfy the equation if we randomly choose the values of a = 2 b =1 c = 2 d = 5 and e =3 and thus we can write the 10-point NHT transformation as G = HT mod 7 which is shown as below:

$$\begin{bmatrix}g(0)\\g(1)\\g(2)\\g(3)\\g(4)\\g(5)\\g(6)\\g(7)\\g(8)\\g(9)\end{bmatrix} = \begin{bmatrix}0 & 2 & 0 & 1 & 0 & 2 & 0 & 5 & 0 & 3\\3 & 0 & 2 & 0 & 1 & 0 & 2 & 0 & 5 & 0\\0 & 3 & 0 & 2 & 0 & 1 & 0 & 2 & 0 & 5\\5 & 0 & 3 & 0 & 2 & 0 & 1 & 0 & 2 & 0\\0 & 5 & 0 & 3 & 0 & 2 & 0 & 1 & 0 & 2\\2 & 0 & 5 & 0 & 3 & 0 & 2 & 0 & 1 & 0\\0 & 2 & 0 & 5 & 0 & 3 & 0 & 2 & 0 & 1\\1 & 0 & 2 & 0 & 5 & 0 & 3 & 0 & 2 & 0\\0 & 1 & 0 & 2 & 0 & 5 & 0 & 3 & 0 & 2\\2 & 0 & 1 & 0 & 2 & 0 & 5 & 0 & 3 & 0\end{bmatrix}\begin{bmatrix}f(0)\\f(1)\\f(2)\\f(3)\\f(4)\\f(5)\\f(6)\\f(7)\\f(8)\\f(9)\end{bmatrix} \bmod 7$$

It is easy to check $NN^T = I \bmod 7$

Table 1. 10-point NHT for 7 (a=2 b=1 c=2 d=5 e=3)

|   | f (n) | g(n) |
|---|---|---|
| 1 | 1,1,1,1,1,1,1,1,1,1 | 6,6,6,6,6,6,6,6,6,6 |
| 2 | 1,1,1,1,0,0,0,0,0,0 | 3,5,5,1,1,0,0,3,3,3 |
| 3 | 0,1,1,1,1,0,0,0,0,0 | 3,3,5,5,1,1,0,0,3,3 |
| 4 | 0,0,1,1,1,1,0,0,0,0 | 3,3,3,5,5,1,1,0,0,3 |
| 5 | 1,1,0,0,0,0,0,0,1,1 | 5,1,1,0,0,3,3,3,3,5 |
| 6 | 1,0,0,1,0,0,1,1,0,0 | 6,5,4,6,4,4,0,4,5,0 |
| 7 | 0,0,1,1,0,0,0,1,0,1 | 2,2,2,3,6,5,8,2,0,1 |
| 8 | 0,1,1,0,0,0,1,1,0,0 | 0,4,5,4,6,0,4,5,4,6 |

Table for different values of a, b, c, d and e, that are different solutions. Note that we have provided two solutions modulo 41.

Table 2. Different solutions for the 10-point NHT

|   | a  | b  | c  | d  | e  | HT mod    | NN$^T$    |
|---|----|----|----|----|----|-----------|-----------|
| 1 | 1  | 4  | 2  | 4  | 3  | HT mod 5  | I mod 5   |
| 2 | 28 | 20 | 6  | 14 | 15 | HT mod 41 | I mod 41  |
| 3 | 1  | 20 | 19 | 35 | 8  | HT mod 41 | I mod 41  |
| 4 | 28 | 55 | 49 | 37 | 13 | HT mod 61 | I mod 61  |
| 5 | 2  | 8  | 3  | 8  | 4  | HT mod 13 | I mod 13  |

Table 3. 10-point NHT for 41 ( a=28 b=20 c=6 d=14 e=15)

|   | f($n$)              | g($n$)                        |
|---|---------------------|-------------------------------|
| 1 | 1,1,1,1,1,1,1,1,1,1 | 1,1,1,1,1,1,1,1,1,1           |
| 2 | 1,1,1,1,0,0,0,0,0,0 | 7,2,2,29,29,20,20,26,26,7     |
| 3 | 0,1,1,1,1,0,0,0,0,0 | 7,7,2,2,29,29,20,20,26,26     |
| 4 | 0,0,1,1,1,1,0,0,0,0 | 25,7,7,2,2,29,29,20,20,26     |
| 5 | 1,1,0,0,0,0,0,0,1,1 | 2,29,29,20,20,26,26,7,7,2     |
| 6 | 1,0,0,1,0,0,1,1,0,0 | 34,21,34,34,35,34,1,35,21,1   |
| 7 | 0,0,1,1,0,0,0,1,0,1 | 8,28,7,15,0,14,21,6,8,20      |
| 8 | 0,1,1,0,0,0,1,1,0,0 | 1,34,21,35,34,1,34,21,35,34   |

We can even assume for large numbers for values of of a =3 b =5 c =10 d =20 and e =40 and 10 point transformation will be G =HT mod 79 and can be checked as NN$^T$ = I mod 79.

**12-point NHT**

Given the first row of the 12-point NHT matrix is 0, a, 0, b, 0, c, 0, d, 0, e, 0, f and is given by the following matrix shown below. By multiplying the 12-point NHT matrix with its transpose we observe that

$$a^2+b^2+c^2+d^2+e^2+f^2$$

is the diagonal element term and the non-diagonal element terms are

(a + e) f + (e + c) d + (c + a) b,
2(ad + be + fc) and
(e + c) a + ec + (b + d) f +bd.

Table 4. NHT for different values of a b c d e f

|   | a  | b  | c  | d  | e  | f  | HT mod     | NN$^T$     |
|---|----|----|----|----|----|----|------------|------------|
| 1 | 1  | 1  | 2  | 4  | 8  | 5  | HT mod 11  | I mod 11   |
| 2 | 33 | 30 | 23 | 9  | 18 | 36 | HT mod 37  | I mod 37   |
| 3 | 2  | 4  | 23 | 16 | 32 | 8  | HT mod 43  | I mod 43   |
| 4 | 2  | 5  | 10 | 7  | 1  | 2  | HT mod 13  | I mod 13   |
| 5 | 26 | 51 | 12 | 6  | 35 | 3  | HT mod 67  | I mod 67   |

Table 5. 12-point NHT for 11 (a=1 b=1 c=2 d=4 e=8 f=5)

|   | f(n)                        | g(n)                                   |
|---|-----------------------------|----------------------------------------|
| 1 | 1,1,1,1,1,1,1,1,1,1,1,1     | 10,10,10,10,10,10,10,10,10,10,10,10    |
| 2 | 1,1,1,1,1,0,0,0,0,0,0,0     | 2,7,6,3,2,6,1,3,6,7,3,4                |
| 3 | 0,1,1,1,1,1,0,0,0,0,0,0     | 4,2,7,6,3,2,6,1,3,6,7,3                |
| 4 | 0,0,1,1,1,1,1,0,0,0,0,0     | 3,4,2,7,6,3,2,6,1,3,6,7                |
| 5 | 1,1,0,0,0,0,0,0,1,1,0,1     | 3,9,6,10,3,5,7,3,4,6,7,9               |
| 6 | 1,0,0,1,0,0,1,1,0,0,1,0     | 5,4,3,2,6,7,9,8,9,10,10,10             |
| 7 | 0,0,1,1,0,0,0,1,0,1,0,1     | 7,1,4,5,1,8,1,4,0,2,5,1                |
| 8 | 0,1,1,0,0,0,1,1,0,0,1,0     | 5,0,7,10,9,0,5,10,7,0,9,10             |

To get a correct NHT matrix we need to select an appropriate matrix in such a way that all of the non diagonal elements in the NN$^T$ product matrix will become zero and only the diagonal elements will remain. There are infinite number of solutions which satisfy the equation we need to choose the values of a=1, b=4, c=8, d=16, e=2, f=11 randomly and we can write the 12-point NHT transformation as G = HT mod 31 which is as follows

$$\begin{bmatrix} g(0) \\ g(1) \\ g(2) \\ g(3) \\ g(4) \\ g(5) \\ g(6) \\ g(7) \\ g(8) \\ g(9) \\ g(10) \\ g(11) \end{bmatrix} = \begin{bmatrix} 0 & 14 & 0 & 28 & 0 & 18 & 0 & 27 & 0 & 23 & 0 & 7 \\ 7 & 0 & 14 & 0 & 28 & 0 & 18 & 0 & 27 & 0 & 23 & 0 \\ 0 & 7 & 0 & 14 & 0 & 28 & 0 & 18 & 0 & 27 & 0 & 23 \\ 23 & 0 & 7 & 0 & 14 & 0 & 28 & 0 & 18 & 0 & 27 & 0 \\ 0 & 23 & 0 & 7 & 0 & 14 & 0 & 28 & 0 & 18 & 0 & 27 \\ 27 & 0 & 23 & 0 & 7 & 0 & 14 & 0 & 28 & 0 & 18 & 0 \\ 0 & 27 & 0 & 23 & 0 & 7 & 0 & 14 & 0 & 28 & 0 & 18 \\ 18 & 0 & 27 & 0 & 23 & 0 & 7 & 0 & 14 & 0 & 28 & 0 \\ 0 & 18 & 0 & 27 & 0 & 23 & 0 & 7 & 0 & 14 & 0 & 28 \\ 28 & 0 & 18 & 0 & 27 & 0 & 23 & 0 & 7 & 0 & 14 & 0 \\ 0 & 28 & 0 & 18 & 0 & 27 & 0 & 23 & 0 & 7 & 0 & 14 \\ 14 & 0 & 28 & 0 & 18 & 0 & 27 & 0 & 23 & 0 & 7 & 0 \end{bmatrix} \begin{bmatrix} f(0) \\ f(1) \\ f(2) \\ f(3) \\ f(4) \\ f(5) \\ f(6) \\ f(7) \\ f(8) \\ f(9) \\ f(10) \\ f(11) \end{bmatrix} \mod 29$$

It is easy to check NN$^T$ = I mod 29.

Table 6. 12-point NHT for 29 (a=14 b=18 c=28 d=27 e=7 f=23)

|   | f $(n)$ | g $(n)$ |
|---|---|---|
| 1 | 1,1,1,1,1,1,1,1,1,1,1,1 | 1,1,1,1,1,1,1,1,1,1,1,1 |
| 2 | 1,1,1,1,1,0,0,0,0,0,0,0 | 3,26,8,15,1,28,5,4,26,15,17,2 |
| 3 | 0,1,1,1,1,1,0,0,0,0,0,0 | 2,3,26,8,15,1,28,5,4,26,15,17 |
| 4 | 0,0,1,1,1,1,1.0,0,0,0,0 | 17,2,3,26,8,15,1,28,5,4,26,15 |
| 5 | 1,1,0,0,0,0,0,0,1,1,0,1 | 15,21,28,6,4,16,15,13,2,12,26,21 |
| 6 | 1,0,0,1,0,0,1,1,0,0,1,0 | 16,29,13,23,12,11,21,11,21,10,6,6 |
| 7 | 0,0,1,1,0,0,0,1,0,1,0,1 | 17,14,18,23,9,7,9,27,24,28,14,18 |
| 8 | 0,1,1,0,0,0,1,1,0,0,1,0 | 12,20,22,10,25,20,12,10,22,20,25,10 |

We can even assume for large numbers for values of of a =78 b =54 c =5 d =10 e=20 and e =40 and 12 point transformation will be G =HT mod 103 and can be checked as $NN^T = I$ mod 103.

**Conclusions**

This paper presented several examples of 10-point and 12-point versions of the recently introduced number theoretic Hilbert (NHT) transforms. Such transforms have applications in signal processing and scrambling. They can be used for generating random sequences with respect to some appropriately chosen modulus. The fact that one has the flexibility of choosing one out of several solutions for a specific modulus enhances the use of these sequences as random sequences.

Further research is required in finding solutions for larger block sizes and developing a theory for the polymorphic solutions for each modulus.